\documentclass[aps,prd,
 amsmath, twocolumn, amssymb,superscriptaddress,
 aps, physrev,
]{revtex4-2}

\usepackage{graphicx}
\usepackage{dcolumn}
\usepackage{bm}
\usepackage{tikz}
\usepackage{derivative}
\usepackage{hyperref}
\usepackage[capitalize]{cleveref}
\usepackage{siunitx}

\usepackage{todonotes} 

\usepackage{placeins}
\usepackage{mathtools}

\usepackage{csquotes}
\usepackage[normalem]{ulem}

\usepackage{my_commands}
\usetikzlibrary{shapes.geometric, arrows.meta, positioning, calc, backgrounds}
\usetikzlibrary{quantikz2}

\usepackage{orcidlink} 

\begin{document}

\title{Prospects and limitations of adaptive variational quantum algorithms for lattice field theories: a case study in 2+1 dimensions}

\author{Nico~Dichter~\orcidlink{0009-0003-0676-8301}}
\email{nico@dichter.eu}
\affiliation{Transdisciplinary Research Area ``Building Blocks of Matter and Fundamental Interactions'' (TRA Matter) and Helmholtz Institute for Radiation and Nuclear Physics (HISKP), University of Bonn, Nussallee 14-16, 53115 Bonn, Germany}

\author{Andrea~Bulgarelli~\orcidlink{0009-0002-2917-6125}}
\affiliation{Transdisciplinary Research Area ``Building Blocks of Matter and Fundamental Interactions'' (TRA Matter) and Helmholtz Institute for Radiation and Nuclear Physics (HISKP), University of Bonn, Nussallee 14-16, 53115 Bonn, Germany}

\author{Lena~Funcke~\orcidlink{0000-0001-5022-9506}}
\affiliation{Transdisciplinary Research Area ``Building Blocks of Matter and Fundamental Interactions'' (TRA Matter) and Helmholtz Institute for Radiation and Nuclear Physics (HISKP), University of Bonn, Nussallee 14-16, 53115 Bonn, Germany}

\author{Stefan~K\"{u}hn~\orcidlink{0000-0001-7693-350X}}
\affiliation{Deutsches Elektronen-Synchrotron DESY, Platanenallee 6, 15738 Zeuthen, Germany}

\date{\today}

\begin{abstract}

    The variational quantum algorithm ADAPT-VQE has previously been used to prepare the vacuum of lattice gauge theories in $1+1D$. In this paper, we study whether it extends to higher dimensions. We identify two general limitations of ADAPT-VQE. First, if the qubit Hamiltonian and the initial state are real, i.e., invariant under qubit-level time-reversal, all real-valued pool operators have vanishing gradients and are never selected. Second, the algorithm can converge to excited eigenstates. We illustrate both limitations, also providing examples from literature, and present a first application of ADAPT-VQE to a lattice field theory in $2+1D$, namely free Wilson fermions, where both obstacles occur. Adding charge conjugation symmetry to the operator pool improves the results, and combining different pools with suitable initial states recovers the ground and first excited states over the full mass range. These findings give practical guidelines for applying ADAPT-VQE to lattice field theories beyond $1+1D$.
    
\end{abstract}

\maketitle

\section{Introduction}

The Hamiltonian formulation of lattice field theories~\cite{Kogut:1974-07} has received increasing attention in recent years, as it gives access to regimes that are hard to reach with conventional Monte Carlo methods. Simulations based on tensor networks or quantum computers do not rely on importance sampling and therefore avoid the sign problem, which affects, for example, theories at finite density, with topological terms, or out of equilibrium. On quantum computers, applications broadly fall into two classes: real-time evolution for the study of non-equilibrium dynamics, and the study of ground-state and low-lying excited-state properties (see, e.g., Refs.~\cite{DiMeglio:2023nsa,Funcke:2023jbq,PRXQuantum.4.027001,bauer2023quantumsimulation,davoudi2026toward} for recent reviews). Both require an efficient way to prepare the ground state, or another state of interest, on the quantum device.

On current noisy hardware~\cite{Preskill:2018-08}, the variational quantum eigensolver (VQE)~\cite{Peruzzo:2014-07, McClean:2016-02, Tilly:2022-09} is a promising tool for this task. A parametrized circuit is executed on the quantum device, and its parameters are optimized classically to minimize the energy. Its applicability is limited by circuit depth~\cite{Koch:2020-08, Urbanek:2021-03, Zhang:2021-08, Cai:2023-10, Algaba:2024-04, Carreras:2026-01}, by barren plateaus~\cite{McClean:2018, Wang:2021, Arrasmith:2021} and local minima~\cite{Anschuetz:2022}, and by the lack of a systematic way to construct the ansatz. ADAPT-VQE~\cite{Grimsley:2019} addresses the last point by growing the circuit iteratively from an operator pool, at each step adding the operator with the largest energy gradient. Its scalable variant, SC-ADAPT-VQE, has prepared the vacuum of the Schwinger model with staggered fermions on more than 100 qubits~\cite{Farrell:2024-04, Farrell2024b}.

This raises the question of whether these algorithms carry over to $2+1D$ and $3+1D$ lattice field theories. This is not only a matter of scale, since beyond $1+1D$, staggered fermions no longer fully remove fermion doubling in the Hamiltonian formulation~\cite{Sharpe:2006-10}. Wilson fermions~\cite{Wilson:1974-08, Rothe:1992, Gattringer:2010} are a common alternative, but they change the symmetries of the qubit Hamiltonian and introduce level crossings in the spectrum~\cite{Golterman:1992-08}. Before ADAPT-VQE can be applied with confidence in higher dimensions, we therefore need to understand when and why this algorithm works, and what its intrinsic limitations are.

In this work, we identify two general constraints of ADAPT-VQE. The first one arises from time-reversal symmetry at the level of the qubits. On qubits, time reversal acts as complex conjugation in the computational basis, so a qubit Hamiltonian is time-reversal symmetric if its matrix elements are real. For such a Hamiltonian and a real initial state, every real-valued pool operator has a vanishing gradient and can never be selected. Importantly, this qubit-level symmetry need not coincide with time-reversal symmetry of the physical system. The second constraint is that every energy eigenstate has vanishing gradients, so ADAPT-VQE can converge prematurely once it reaches an excited state. We further analyze how incorporating the symmetries of the theory into the operator pool and the initial state, an established technique~\cite{Bertels:2022oga, Farrell:2024-04, Farrell2024b, Shirali:2025-10}, affects the convergence of ADAPT-VQE.

After illustrating these constraints with examples from the literature, namely the Schwinger model in $1+1D$~\cite{Farrell:2024-04} and supersymmetric quantum mechanics~\cite{Kerfoot:2025mvx}, we present a first application of ADAPT-VQE to a lattice field theory in $2+1D$: free Wilson fermions on a $2\times 2$ lattice. This exactly solvable system~\cite{Kivlichan:2018} is a controlled testbed. Its qubit Hamiltonian is real even though physical time-reversal symmetry is broken, and ADAPT-VQE prematurely converges at excited states, so both obstacles occur. Operator pools that respect charge conjugation improve the results, but no single setup reaches the ground state at every mass. Nevertheless, by combining different pools with suitably chosen initial states, we reconstruct both the ground and the first excited state over the full mass range. Our analysis thus provides guidelines for applying ADAPT-VQE to lattice field theories beyond $1+1D$.

The remainder of the paper is organized as follows. Section~\ref{sec:algorithms} reviews VQE and ADAPT-VQE. Section~\ref{sec:constraints} derives the algorithmic constraints and revisits examples from the literature. Section~\ref{sec:wilson-2+1d} presents the $2+1D$ case study, before finally concluding in Section~\ref{sec:conclusions}.

\section{Algorithms}
\label{sec:algorithms}

Here, we briefly review VQE and ADAPT-VQE and introduce the notation used throughout the paper.

\subsection{The Variational Quantum Eigensolver}\label{sec:vqe-alg}
VQE, firstly introduced by \textcite{Peruzzo:2014-07}, is a hybrid quantum-classical algorithm to optimize a parametrized state by minimizing the expectation value of a given cost observable $O$ that can be efficiently measured on a quantum computer.
Here we will always choose $O$ to be the Hamiltonian of the system of interest,
possibly with the addition of penalty terms.

VQE relies on two ingredients: an initial state $\ket{\psi_0}$ that is easy to prepare on a quantum computer, and a parametrized circuit $U(\vec{\theta})$, whose structure may be informed by prior knowledge of the system. The algorithm proceeds as follows:

\begin{enumerate}
    \item Initialize a vector of parameters $\Vec{\theta}^{(0)}$.
    \label{enum:vqe-init-params}
    \item Prepare the state $\ket{\psi\brac*{\vec{\theta}^{(n)}}} = U\brac*{\vec{\theta}^{(n)}}\Big|\psi_0\Big>$ on a quantum computer. \label{enum:vqe-prep-state}
    \item Measure the expectation value of the cost observable $O$ on a quantum computer: \label{enum:vqe-meas-cost}
    \begin{equation*}
        f_c\brac*{\vec{\theta}^{(n)}}=\langle O \rangle_{\vec{\theta}^{(n)}}=\bra{\psi\brac*{\vec{\theta}^{(n)}}} O \ket{\psi\brac*{\vec{\theta}^{(n)}}}.
    \end{equation*}
    \item Pass the cost function $f_c\brac*{\vec{\theta}^{(n)}}$ to a classical optimizer, which proposes updated parameters $\vec{\theta}^{(n+1)}$. If the optimizer has converged, terminate and return $\ket{\psi\brac*{\vec{\theta}^{(n)}}}$ as the final state; otherwise, set $n \rightarrow n+1$ and return to step~\ref{enum:vqe-prep-state}. \label{enum:vqe-optimize}
\end{enumerate}
Steps~\ref{enum:vqe-prep-state} and~\ref{enum:vqe-meas-cost} are performed on the quantum device, whose $N$-qubit register naturally spans a $2^N$-dimensional Hilbert space that is, in general, exponentially costly to represent classically. The classical optimization in step~\ref{enum:vqe-optimize} can be performed with gradient-free optimizers such as COBYLA~\cite{Powell:1994} and COBYQA~\cite{Ragonneau-phd-thesis}, or gradient-based ones such as SLSQP~\cite{Kraft:1988}. Independently of this choice, VQE can suffer from barren plateaus, i.e.\ large regions in parameter space where the gradients of the cost function are vanishing exponentially with system size~\cite{McClean:2018,Wang:2021}, which has been shown to be equivalent to the cost function being exponentially concentrated in narrow gorges~\cite{Arrasmith:2021}. In addition, even in the absence of barren plateaus, the scarcity of good local minima in the cost landscape can render the optimization difficult~\cite{Anschuetz:2022}.

A further challenge lies in the design of the circuit itself, as there is no systematically improvable procedure for constructing a variational ansatz. This motivates the search for a variational algorithm that does not require the circuit to be fixed \textit{a priori}.

\subsection{ADAPT-VQE}\label{subsec:adapt-vqe-alg}

ADAPT-VQE, first introduced by \textcite{Grimsley:2019}, extends VQE by replacing the fixed parametrized circuit $U(\vec{\theta})$ with an adaptive routine that constructs the circuit iteratively, guided by the gradients of a given cost function.

Like VQE, ADAPT-VQE relies on two main ingredients. The first is again an initial state $\ket{\psi_0}$ that is easy to prepare on a quantum computer. The parametrized circuit is instead replaced by a pool of Hermitian operators $P=\set{A_i\middle|i\in \set{1,\dots, N_{\text{Pool}}}}$, whose composition may be informed by prior knowledge of the system. Given a cost observable $O$ the algorithm then proceeds as follows:

\begin{figure*}
    \centering
    \includegraphics[width=\linewidth]{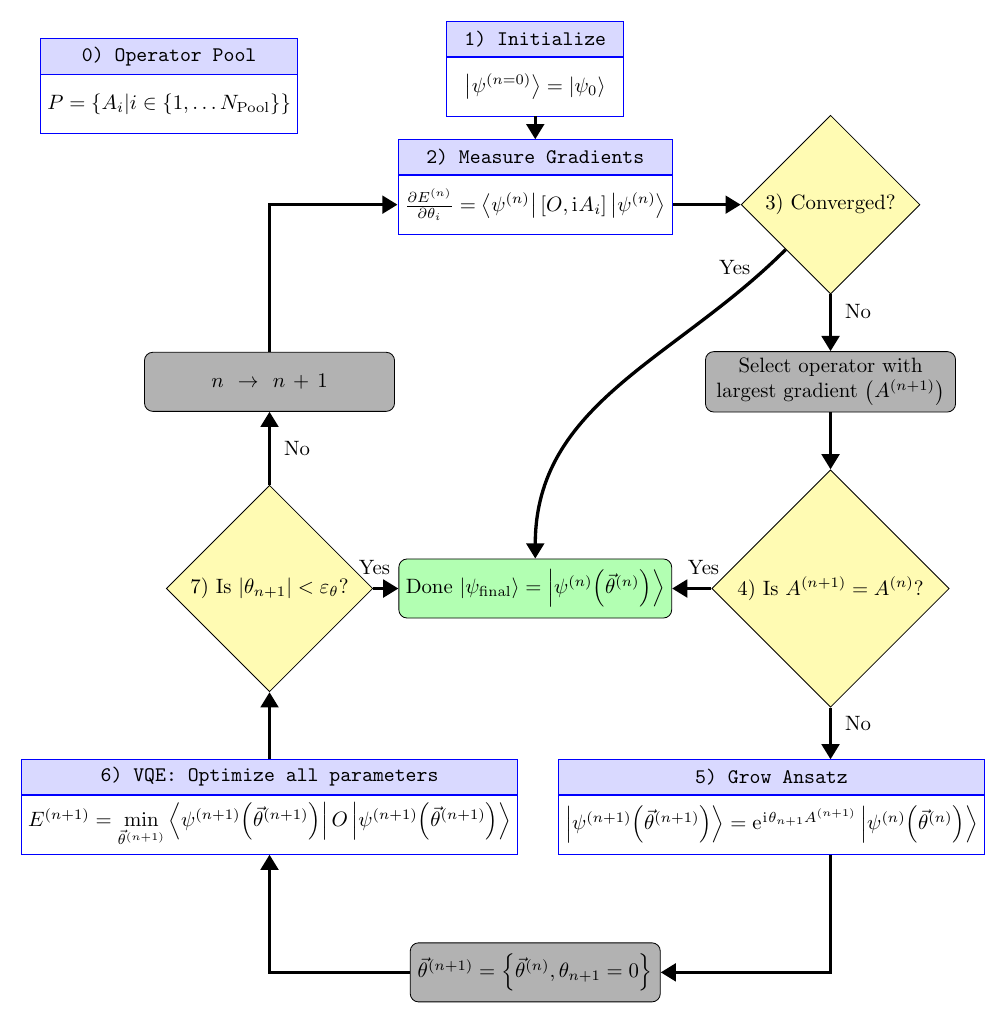}
    \caption[ADAPT-VQE flow chart]{A flow chart describing the main steps of the ADAPT-VQE algorithm. For a detailed description and an explanation of the notation see~\autoref{subsec:adapt-vqe-alg}. Note that the original ADAPT-VQE algorithm~\cite{Grimsley:2019} only included the exit option of step~\ref{enum:avqe-exit-1}, the exit option of step~\ref{enum:avqe-exit-3} was introduced in Ref.~\cite{Farrell:2024-04}, and the exit option of step~\ref{enum:avqe-exit-2} is a modification introduced in this work.}
    \label{fig:adapt_vqe_flow_chart}
\end{figure*}

\begin{enumerate}
    \item Initialize the state to $\ket{\psi^{(n=0)}\brac*{\vec{\theta}^{(n=0)}}}=\ket{\psi_0}$. The parameter vector $\vec{\theta}^{(n=0)}$ is empty.
    \item Measure the gradients in the \emph{new} $\theta_{n+1}$ direction of the cost observable landscape
    for every operator $A_i$ in the pool: \label{enum:avqe-grad}
    \begin{align}
        \pdv{E^{(n)}}{\theta_i}&\coloneqq \pdv{}{\theta_i}\bigg(\bra{\psi^{(n)}}\euler^{-\iu \theta_i A_i}O\euler^{\iu \theta_i A_i}\ket{\psi^{(n)}}\bigg\vert_{\theta_i=0}\nonumber\\
        &=\bra{\psi^{(n)}}\comm{O}{\iu A_i}\ket{\psi^{(n)}}. \label{eq:num-avqe-grad}
    \end{align}
    \item If the magnitude of the vector of gradients 
    \begin{align}
            \sqrt{ \sum\limits_{i=1}^{N_{\text{Pool}}}\brac*{\pdv{E^{(n)}}{\theta_i}}^2},
    \end{align}
    is smaller than a predefined threshold $\varepsilon$, exit with the final state given by $\ket{\psi^{(n)}\brac*{\vec{\theta}^{(n)}}}$. Otherwise, select the operator with the largest associated gradient in magnitude, name it $A^{(n+1)}$. \label{enum:avqe-exit-1}
    \item If the selected operator $A^{(n+1)}$ is the same operator as the one selected in the previous iteration $A^{(n)}$, exit with the final state given by $\ket{\psi^{(n)}\brac*{\vec{\theta}^{(n)}}}$. \label{enum:avqe-exit-2}
    \item Grow the ansatz state with the additional unitary operator $\euler^{\iu \theta_{n+1} A^{(n+1)}}$:
    \begin{align}
        &\ket{\psi^{(n+1)}\brac*{\Vec{\theta}^{(n+1)}}}\nonumber\\
        &=\euler^{\iu \theta_{n+1} A^{(n+1)}}\ket{\psi^{(n)}\brac*{\Vec{\theta}^{(n)}}}. \label{eq:avqe-ansatz-grow}
    \end{align}
    This adds a \emph{new} parameter $\theta_{n+1}$ to the ansatz which is initialized to $\theta_{n+1}=0$. \label{enum:avqe-grow}
    \item Using the initial parameter vector of $\Vec{\theta}^{(n+1)}=\set{\Vec{\theta}^{(n)}, \theta_{n+1}=0}$, run a VQE optimization subroutine. This routine starts at step~\ref{enum:vqe-prep-state} of the VQE algorithm and optimizes all parameters.
    \item If 
    $\abs{\theta_{n+1}}$ is smaller than a predefined threshold $\varepsilon_\theta$, exit with the final state given by $\ket{\psi^{(n)}\brac*{\vec{\theta}^{(n)}}}$. In this instance, by abuse of notation, $\vec{\theta}^{(n)}$ denotes the first $n$ components of the optimized parameter vector $\vec{\theta}^{(n+1)}$.\label{enum:avqe-exit-3}
    \item Go back to step~\ref{enum:avqe-grad} with $n\rightarrow n+1$.
\end{enumerate}

A visual representation of these steps is shown in \autoref{fig:adapt_vqe_flow_chart}. The newly added parameter in step~\ref{enum:avqe-grow} is initialized to zero, since the gradients in step~\ref{enum:avqe-grad} are evaluated at this value. The original ADAPT-VQE algorithm~\cite{Grimsley:2019} only included the exit option of step~\ref{enum:avqe-exit-1}, the exit option of step~\ref{enum:avqe-exit-3} was introduced in Ref.~\cite{Farrell:2024-04}, and the exit option of step~\ref{enum:avqe-exit-2} is a modification introduced in this work. It is motivated by the fact that for $A^{(n+1)}=A^{(n)}$ the newly added parameter is redundant, since
\begin{equation}
    \euler^{\iu \theta_{n+1} A^{(n)}}\euler^{\iu \theta_{n} A^{(n)}}=\euler^{\iu \brac*{\theta_{n}+\theta_{n+1}} A^{(n)}}.
\end{equation}

Notice that the gradient threshold $\varepsilon$, the parameter threshold $\varepsilon_\theta$, and the target accuracy $\varepsilon_{\text{opt}}$ of the optimizer in the VQE subroutine should be tuned consistently. Specifically, if $\varepsilon_{\text{opt}}$ is set too large relative to $\varepsilon$, the same operator may be selected in two consecutive iterations. Similarly, if $\varepsilon_{\text{opt}}$ is set too large relative to $\varepsilon_\theta$, the exit condition of step~\ref{enum:avqe-exit-3} may fail to trigger, again leading to consecutive selections of the same operator.

\section{Algorithmic constraints of ADAPT-VQE}
\label{sec:constraints}

\subsection{Time reversal symmetry}\label{sec:time-reversal-symmetry}

In this section, we show that time-reversal symmetry at the qubit level, i.e.\ after mapping the original theory to qubits, strongly constrains which operators ADAPT-VQE can select. On qubits, time reversal acts as complex conjugation in the computational basis; a qubit Hamiltonian is therefore time-reversal symmetric if its matrix elements in this basis are real. For such a Hamiltonian and a real-valued wave function, we find that
\begin{itemize}
    \item the gradient in Eq.~\eqref{eq:num-avqe-grad} is independent of the real part of the components of the pool operator $A$,
    \item the gradient vanishes identically if all components of $A$ are real valued, such that these operators are never selected, and ADAPT-VQE exits in the first iteration if the pool contains only such operators,
    \item if the selected operators are purely imaginary, the wave function remains real after each iteration, so ADAPT-VQE is constrained to real-valued wave functions. This does not limit the search for energy eigenstates, since the eigenvectors of a real Hamiltonian can always be chosen to be real valued.
\end{itemize}
Note that time reversal symmetry of the qubit Hamiltonian does not need to coincide with time reversal symmetry of the original physical system. In particular for field theories, time-reversal symmetry in general acts nontrivially on the spinor components of the fields, see~\autoref{sec:d-symmetries}. The role of time reversal symmetry on ADAPT-VQE was already briefly mentioned in Ref.~\cite{Tang:2021-04}; here we derive it explicitly and discuss its consequences in the context of lattice field theories.

In what follows we combine the string of binary indices 
\[(i_0,\dots, i_{N-1})~\text{where}~\forall j\in \set{0,\dots, N-1}:~i_j\in \set{0,1}\] of a general $N$ qubit state into a single index $i\in\set{0,\dots,2^{N}-1}$. With this we can write 
\begin{align}
    \ket{\psi}
    &=\sum\limits_{i=0}^{2^N-1} c_{i}\ket{i}. \label{eq:num-psi-multi}
\end{align}
A Hamiltonian acting on $N$ qubits then becomes
\begin{equation}
    H= \sum\limits_{i,j=0}^{2^N-1} h_{ij}\ket{i}\bra{j}, \label{eq:num-h-multi}
\end{equation}
and similarly any hermitian operator $A$ acting on $N$ qubits
\begin{equation}
    A= \sum\limits_{i,j=0}^{2^N-1} a_{ij}\ket{i}\bra{j}. \label{eq:num-a-multi}
\end{equation}
Here $c_{i}, h_{ij}, a_{ij}\in \complexnumbers$ and $\ket{i}=\ket{i_0,\dots, i_{N-1}}$ denote $N$ qubit computational basis states.

To prove the statements summarized above, we decompose the gradient in Eq.~\eqref{eq:num-avqe-grad} into the real and imaginary parts of the quantities appearing in it
\begin{align}
    a_{ij}&=a_{ij}^R+\iu a_{ij}^I\nonumber\\
    \text{where}\quad a_{ij}^R&=a_{ji}^R\quad, \quad a_{ij}^I=-a_{ji}^I \label{eq:num-a-r-i-decomp}\\
    \text{and}\quad a_{ij}^R,a_{ij}^I&\in \realnumbers. \nonumber
\end{align}
Similarly for the Hamiltonian:
\begin{align}
    h_{ij}&=h_{ij}^R+\iu h_{ij}^I. \label{eq:num-h-r-i-decomp}
\end{align}
With this we can write the gradient used in ADAPT-VQE as
\begin{align}
    \pdv{E}{\theta}&\overset{\ref{eq:num-avqe-grad}}{=}\bra{\psi}\comm{H}{\iu A}\ket{\psi}\nonumber\\
    &\overset{\ref{eq:num-psi-multi}}{\overset{\ref{eq:num-h-multi}}{\overset{\ref{eq:num-a-multi}}{=}}} \iu c_i^*\brac*{h_{ij} a_{jk}- a_{ij}h_{jk}}c_k\nonumber\\
    &\overset{\ref{eq:num-a-r-i-decomp}}{\overset{\ref{eq:num-h-r-i-decomp}}{=}}
     c_i^*\bigg[\iu\brac*{h_{ij}^R a_{jk}^R- a_{ij}^Rh_{jk}^R}- \iu\brac*{h_{ij}^I a_{jk}^I- a_{ij}^Ih_{jk}^I}\nonumber \\ 
     &\qquad\quad-\brac*{h_{ij}^R a_{jk}^I- a_{ij}^Ih_{jk}^R}-\brac*{h_{ij}^I a_{jk}^R- a_{ij}^Rh_{jk}^I}\bigg]c_k \nonumber \\
     &= c_i^*\left[\iu\alpha_{ik}- \iu\beta_{ik} -\varphi_{ik}-\xi_{ik}\right]c_k \label{eq:num-avqe-grad-r-i-decomp-1}
\end{align}
where the summation over the multi indices is implicit from now on. And in the last step we introduced the following components:
\begin{align}
    \alpha_{ik} &= h_{ij}^R a_{jk}^R- a_{ij}^Rh_{jk}^R = - \brac*{a_{ij}^Rh_{jk}^R - h_{ij}^R a_{jk}^R}\nonumber \\
    &\overset{\ref{eq:num-a-r-i-decomp}}{\overset{\ref{eq:num-h-r-i-decomp}}{=}} - \brac*{a_{ji}^Rh_{kj}^R - h_{ji}^R a_{kj}^R}=-\alpha_{ki} ,\label{eq:num-avqe-alpha-1}\\
    \beta_{ik} &= h_{ij}^I a_{jk}^I- a_{ij}^Ih_{jk}^I = - \brac*{a_{ij}^Ih_{jk}^I - h_{ij}^I a_{jk}^I}\nonumber \\
    &\overset{\ref{eq:num-a-r-i-decomp}}{\overset{\ref{eq:num-h-r-i-decomp}}{=}} - \brac*{a_{ji}^Ih_{kj}^I - h_{ji}^I a_{kj}^I}=-\beta_{ki} ,\label{eq:num-avqe-beta-1}\\
    \varphi_{ik} &= h_{ij}^R a_{jk}^I- a_{ij}^Ih_{jk}^R = - \brac*{a_{ij}^Ih_{jk}^R - h_{ij}^R a_{jk}^I}\nonumber \\
    &\overset{\ref{eq:num-a-r-i-decomp}}{\overset{\ref{eq:num-h-r-i-decomp}}{=}} \brac*{a_{ji}^Ih_{kj}^R - h_{ji}^R a_{kj}^I}=\varphi_{ki} ,\label{eq:num-avqe-varphi-1}\\
    \xi_{ik} &= h_{ij}^I a_{jk}^R- a_{ij}^Rh_{jk}^I = - \brac*{a_{ij}^Rh_{jk}^I - h_{ij}^I a_{jk}^R}\nonumber \\
    &\overset{\ref{eq:num-a-r-i-decomp}}{\overset{\ref{eq:num-h-r-i-decomp}}{=}}\brac*{a_{ji}^Rh_{kj}^I - h_{ji}^I a_{kj}^R}=\xi_{ki}. \label{eq:num-avqe-xi-1}
\end{align}

We will now assume that the wave-function in the computational basis is real valued up to an irrelevant global phase, namely $c_i \in \realnumbers$, then we can introduce the symmetric components 
\begin{equation}
    c_{ik}=c_ic_k=c_i^*c_k=c_{ki}. \label{eq:num-avqe-symmetric-wf-components}
\end{equation}
With these Eq.~\eqref{eq:num-avqe-grad-r-i-decomp-1} can be simplified as follows:
\begin{align}
    \begin{aligned}
        \pdv{E}{\theta}
        &\overset{\ref{eq:num-avqe-grad-r-i-decomp-1}}{=}
         c_i^*\left[\iu\alpha_{ik}- \iu\beta_{ik} -\varphi_{ik}-\xi_{ik}\right]c_k\\
         &= c_{ik}\left[\iu\alpha_{ik}- \iu\beta_{ik} -\varphi_{ik}-\xi_{ik}\right]\\
         &\overset{\ref{eq:num-avqe-alpha-1}}{\overset{\ref{eq:num-avqe-beta-1}}{\overset{\ref{eq:num-avqe-symmetric-wf-components}}{=}}} - c_{ik}\left[ \varphi_{ik}+\xi_{ik}\right]\\
         &\overset{\ref{eq:num-avqe-varphi-1}}{\overset{\ref{eq:num-avqe-xi-1}}{=}}
         - c_{ik}\left[ h_{ij}^R a_{jk}^I- a_{ij}^Ih_{jk}^R+h_{ij}^I a_{jk}^R- a_{ij}^Rh_{jk}^I\right].
     \end{aligned}
\end{align}

For a time-reversal-symmetric qubit Hamiltonian, $h_{ij}^I=0$, this reduces to
\begin{equation}
    \pdv{E}{\theta} = - c_{ik}\left[ h_{ij}^R a_{jk}^I- a_{ij}^Ih_{jk}^R\right],
\end{equation}
which depends only on the imaginary part of $A$ and thus proves the first two statements. The third statement follows since a purely imaginary operator, $a_{ij}^R=0$, generates a real orthogonal gate $\euler^{\iu\theta A}$, such that the wave function remains real after each extension of the ansatz according to Eq.~\eqref{eq:avqe-ansatz-grow}, and the argument can be applied iteratively.

\subsection{Premature convergence at excited states}\label{sec:higher-excited-states}

Since energy eigenstates are stationary points of the energy functional, ADAPT-VQE can terminate prematurely in an excited state. While this behavior has been observed numerically in quantum chemistry applications~\cite{Grimsley:2022azc}, we here discuss this obstacle and show that it becomes particularly relevant for (lattice field) theories with level crossings, where the ground state changes its character as a function of the parameters of the theory, as we observe for Wilson fermions in \autoref{sec:wilson-2+1d}.\\ To see why ADAPT-VQE terminates in such cases, note that the gradients vanish for every energy eigenstate $\ket{\psi}$, $H\ket{\psi}=E_\psi\ket{\psi}$. Inserting this into Eq.~\eqref{eq:num-avqe-grad} yields
\begin{equation}
    \pdv{E}{\theta_i}=\bra{\psi}\comm{H}{\iu A_i}\ket{\psi}=\iu E_\psi\bra{\psi}\brac*{A_i-A_i}\ket{\psi}=0
\end{equation}
for all pool operators $A_i$.
Consequently, once an energy eigenstate is prepared with sufficient fidelity, all gradients in Eq.~\eqref{eq:num-avqe-grad} approximately vanish and ADAPT-VQE converges prematurely. This limitation might seem unavoidable but it can be mitigated and in some respect even circumvented by employing a non local selection criterion as discussed in~\autoref{sec:conclusions}.

\subsection{The role of symmetries}\label{sec:symmetries}

Incorporating symmetries into ADAPT-VQE by restricting the operator pool to symmetry-preserving operators and choosing the initial state in a definite symmetry sector is an established technique~\cite{Bertels:2022oga, Farrell:2024-04,Farrell2024b,Shirali:2025-10}. Here, we briefly review this scheme for completeness and analyze how it affects the convergence of ADAPT-VQE, namely by avoiding energetically flat directions and by reducing the dimension of the reachable Hilbert space and the size of the operator pool.

A symmetry can be represented by an (anti-)unitary operator $S$ on the Hilbert space that commutes with the Hamiltonian.
Such a symmetry can be incorporated into the operator pool by imposing
\begin{align}
    \forall A_i \in P:\, \euler^{\iu\theta A_i}&\overset{!}{=}S \euler^{\iu\theta A_i}S^{-1}=\euler^{\theta S \iu A_i S^{-1}}.
\end{align}

Which is equivalent to imposing 
\begin{equation}
    \comm{A_i}{S}=0 \label{eq:comm-general-symmetry}
\end{equation}
if $S$ is unitary and
\begin{equation}
    \acomm{A_i}{S}=0
\end{equation}
if $S$ is anti-unitary.

By then choosing the initial state $\ket{\psi_0}$ to be an eigenstate of $S$
\begin{equation}
    S\ket{\psi_0}=\alpha_{S,0}\ket{\psi_0},\label{eq:eigenstate-general-symmetry}
\end{equation}
we are guaranteed~\footnote{This discussion assumes negligible noise levels. For real noisy hardware additional penalty terms have to be added to the cost operator, as shown in \autoref{sec:numerics-setup} for the charge operator $Q$.
} 
to obtain an eigenstate of $S$ as the final state using ADAPT-VQE with the same eigenvalue as the initial state $\ket{\psi_0}$:
\begin{align}
    \begin{aligned}
        S\ket{\psi^{(n)}\left(\Vec{\theta^{(n)}}\right)}
        &\overset{\ref{eq:avqe-ansatz-grow}}{=}S\euler^{\iu \theta_n A^{(n)}}\dots \euler^{\iu \theta_1 A^{(1)}} \ket{\psi_0}\\
        &=\euler^{\iu \theta_n A^{(n)}}\dots \euler^{\iu \theta_1 A^{(1)}} S\ket{\psi_0}\\
        &\overset{\ref{eq:eigenstate-general-symmetry}}{=} \euler^{\iu \theta_n A^{(n)}}\dots \euler^{\iu \theta_1 A^{(1)}} \alpha_{S,0}\ket{\psi_0}\\
        &\overset{\ref{eq:avqe-ansatz-grow}}{=} \alpha_{S,0}\ket{\psi^{(n)}\left(\Vec{\theta^{(n)}}\right)}.
    \end{aligned}
\end{align}

We now discuss how this scheme affects the convergence behaviour of ADAPT-VQE:

\begin{itemize}
    \item \textit{Avoidance of flat directions.} Consider two degenerate energy eigenstates satisfying $S\ket{\psi_{s_1}} = s_1\ket{\psi_{s_1}}$ and $S\ket{\psi_{s_2}} = s_2\ket{\psi_{s_2}}$ with $s_1\neq s_2$. If the symmetry $S$ is not incorporated into the operator pool, ADAPT-VQE can converge to an arbitrary superposition of $\ket{\psi_{s_1}}$ and $\ket{\psi_{s_2}}$. If $S$ is incorporated, ADAPT-VQE reproduces either $\ket{\psi_{s_1}}$ or $\ket{\psi_{s_2}}$, depending on the choice of the initial state $\ket{\psi_0}$, thereby avoiding a search along the energetically flat direction connecting the two states.

    \item \textit{Reduced search space.} The dimension of the relevant Hilbert space is reduced. At the same time, the operator pool shrinks, since it can be restricted to operators compatible with $S$.

    \item \textit{Improved accuracy.} Empirically, by incorporating charge conjugation symmetry into the ansatz, we find an improvement of the reconstruction of energy eigenstates, as we show in~\autoref{sec:wilson-2+1d}.
\end{itemize}

\subsection{Examples from the literature}
\subsubsection{Explaining the performance of ADAPT-VQE in the Schwinger model}
The case of a real valued  Hamiltonian and wave function as considered in \autoref{sec:time-reversal-symmetry} might seem too specific. Yet one of the most successful applications of SC-ADAPT-VQE to the Schwinger model by~\textcite{Farrell:2024-04} falls exactly into this category.

The problem described in~\autoref{sec:time-reversal-symmetry} was avoided in that work because time reversal (\timereversal) symmetry on the level of the qubits was imposed on the operator pool. Specifically
\begin{align*}
    \forall A_i \in P:\, \euler^{\iu\theta A_i}&\overset{!}{=}\timereversal \euler^{\iu\theta A_i}\timereversal^{-1}\\
    &=\euler^{-\iu\theta \timereversal A_i \timereversal^{-1}}.
\end{align*}
Since, when acting on qubits, the time reversal symmetry is just a complex conjugation, this is equivalent to the condition
\begin{equation*}
    A_i=-A_i^*.
\end{equation*}
Imposing this condition restricts the pool to purely imaginary operators, which avoids the problem identified in \autoref{sec:time-reversal-symmetry}. As that section shows, however, incorporating time-reversal symmetry is not merely an option for improving efficiency, as are the symmetries discussed in~\autoref{sec:symmetries}, but a \emph{necessity} for ADAPT-VQE to operate at all.

In addition, the charge-conjugation and parity (\chargeconjugation\parity) symmetries of the physical system were imposed on the operator pool. As discussed in \autoref{sec:symmetries}, this improves the convergence behaviour of ADAPT-VQE in several ways.

\subsubsection{Supersymmetric Quantum Mechanics}
An additional example of the limitations discussed in~\autoref{sec:time-reversal-symmetry} can be found in recent work by~\textcite{Kerfoot:2025mvx}. There they consider a real valued Hamiltonian and wave function and propose an operator pool consisting of $\{ R_\mathbf{Y}, R_\mathbf{Z}, cR_\mathbf{Y} \}$. Since $R_\mathbf{Z}$ is a real operator, \autoref{sec:time-reversal-symmetry} implies that its gradient vanishes identically for any real state. It can therefore never be selected and only adds to the measurement cost of each ADAPT-VQE iteration. This shows that the obstacle outlined in~\autoref{sec:time-reversal-symmetry} also arises outside of lattice field theory.

Notice that, strictly speaking, the algorithm used in~\cite{Kerfoot:2025mvx} is not equivalent to ADAPT-VQE, as controlled rotations are also included in the pool, and the latter cannot be written as an exponential of an hermitian operator. Nonetheless, our result extends to this case. 

\section{A case study in 2\,+\,1\textit{D}} \label{sec:wilson-2+1d}

We now turn to a first case study in $2+1D$, namely free massive Wilson fermions~\cite{Wilson:1974-08}. To the best of our knowledge, this is the first application of ADAPT-VQE to a lattice field theory in $2+1D$. Since the Hamiltonian of free fermions is quadratic, its ground state is a Slater determinant, which can be prepared exactly with quantum circuits of known depth~\cite{Kivlichan:2018}. Thus, we do not aim to compete with such established non-variational methods, but use the free theory as a testbed: its spectrum, including the level crossings at $m+2r\in\set{-2,0,2}$, is known, and its qubit Hamiltonian exhibits both obstacles discussed in \autoref{sec:time-reversal-symmetry} and \autoref{sec:higher-excited-states}.

\subsection{Theoretical setup}\label{sec:wilson-hamiltonian}

The lattice Hamiltonian considered hereby reads 
\begin{align}
    &H_{\text{Free}}\label{eq:num-h-free-tot-bc}\\
    &=\sum\limits_{n\in\Lambda}\Bigg[-\dfrac{1}{2}\sum\limits_{j=1}^{2}\left\{\Psibar(n)\left(\iu\gamma^j+r\right)\Psi \left(n+\widehat{j}\right)+\hermitc\right\}\nonumber\\
    &\qquad\qquad+\Psibar(n)\left(m+2 r\right)\Psi(n)\Bigg],\nonumber
\end{align}
where $m\in \realnumbers$ is the mass parameter,  $r\in \realnumbers$ is the Wilson parameter,
\begin{equation}
    \Lambda=\set{(n_x, n_y)\big\vert n_i\in \set{0,\dots, N_i-1}}\label{eq:num-free-naive-lambda}
\end{equation}
denotes the introduced \emph{periodic} lattice and $\set{N_x,N_y}$ are the spatial extents of $\Lambda$ in units of the lattice spacing $a$. 
$\widehat{j}$ denotes the unit vector in the $j^{\text{th}}$ spatial direction.

In $2+1D$ $\Psi(n)$ is a two-component spinor.
We choose to map the upper component to $m_x=2n_x$ and the lower component to $m_x=2n_x+1$:
\begin{equation}
    \Psi(n)=\Psi(n_x,n_y)\eqqcolon\begin{pmatrix}
        \Phi(2n_x,n_y)\\
        \Phi(2n_x+1,n_y)
    \end{pmatrix},\label{eq:num-psi-phi}
\end{equation}
such that $\Phi(m_x,n_y)$ are fermionic modes defined on $m_x\in \set{0,\dots, M_x-1}$, $n_y\in \set{0,\dots, N_y-1}$, where we defined $M_x=2N_x$.

Equation~\eqref{eq:num-h-free-tot-bc} can then be expressed in terms of $\Phi(m_x,n_y)$ (see\hyperref[sec:ham-ferm-modes]{App.~\ref*{sec:ham-ferm-modes}} for details).

\subsubsection{Mapping fermions to qubits}\label{sec:map-fermions-qubits}

To simulate the theory on a quantum computer, we map the fermionic modes $\Phi(m_x,n_y)$ introduced in Eq.~\eqref{eq:num-psi-phi} to qubits, using the Jordan-Wigner transformation (JWT)~\cite{Jordan:1928-09}. We briefly review this mapping here, since it involves two choices that become relevant later: an ordering of the fermionic modes on the two-dimensional lattice, and a phase convention for each mode. The latter determines whether the resulting qubit Hamiltonian has real or complex components, and thus whether the obstacle of~\autoref{sec:time-reversal-symmetry} occurs.

\begin{figure}
    \centering
    \includegraphics[width=\linewidth]{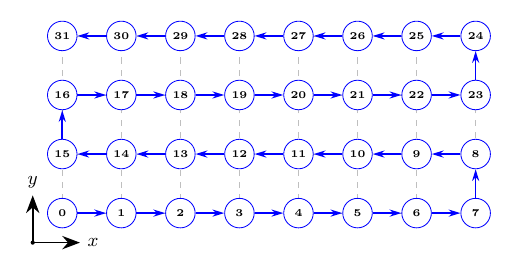}
    \caption[Snake ordering for fermionic modes ($8\times 4$)]{Ordering snake for fermionic modes on a $(M_x=8)\times (4=N_y)$ lattice.}
    \label{fig:j-snake-general}
\end{figure}

For the JWT, we need to fix a linear ordering in the fermionic modes $\Phi(m_x,n_y)$. Since these modes are already defined on a two dimensional lattice with side lengths $M_x=2N_x$ and $N_y$, we can choose a snake ordering as shown in \autoref{fig:j-snake-general}.
The ordering index $j$ can then be obtained from the lattice coordinates $(m_x, n_y)$ as follows:
\begin{equation}
    j(m_x,n_y)\coloneqq \begin{dcases}
        M_x n_y +m_x &n_y \text{ even},\\
        M_x (n_y+1) - (m_x+1) &n_y \text{ odd}.
    \end{dcases} \label{eq:qc-j-snake}
\end{equation}
The details of this transformation are shown in \hyperref[sec:jwt-a]{App.~\ref*{sec:jwt-a}}. The result obtained there is that the operators are mapped as
\begin{align}
    \Phi(m_x,n_y)&\eqqcolon\Phi(j(m_x,n_y))=\Phi_j\\\Phi_j &=\brac*{\prod\limits_{n=0}^{j-1} \pz\qubit{n}}\brac*{\sigma^+}\qubit{j},\label{eq:qc-jwt-phi}\\
    \Phi_j^\dagger &=\brac*{\prod\limits_{n=0}^{j-1} \pz\qubit{n}}\brac*{\sigma^-}\qubit{j},\label{eq:qc-jwt-phi-dagger}
\end{align}
where $\sigma^\pm=\frac{1}{2}\brac*{\px\pm\iu\py}$, \px\,,\py\, and \pz\, denote the Pauli matrices
and a general state
{
\newcommand*\fermionvac{\omega}
\begin{align}
    \ket{\psi} &= \sum\limits_{i_0,\dots, i_{N-1}=0}^1 C_{i_0,\dots, i_{N-1}}\underbrace{\left(\Phi_0^\dagger\right)^{i_0} \dots \left(\Phi_{N-1}^\dagger\right)^{i_{N-1}} \ket{\fermionvac}}_{\eqqcolon \ket{i_0,\dots, i_{N-1}}}\nonumber \\
    &= \sum\limits_{i_0,\dots, i_{N-1}=0}^1 C_{i_0,\dots, i_{N-1}}\ket{i_0,\dots, i_{N-1}},\label{eq:qc-jwt-general-state}
\end{align}
is mapped to the corresponding qubit state where we have introduced $N\coloneqq M_xN_y$ and defined the reference state $\ket{\fermionvac}=\ket{0,\dots,0}$ to satisfy:
\begin{equation}
    \forall j \in \set{0,\dots, N-1}:\quad \Phi_j\ket{\fermionvac}=0.
    \label{eq:qc-jwt-reference-state}
\end{equation}
}

With the fermionic modes $\Phi_j$ faithfully mapped to qubits, we can now express $H_{\text{Free}}$ in terms of operators acting on qubit states. Inserting Eqs.~\eqref{eq:qc-jwt-phi} and~\eqref{eq:qc-jwt-phi-dagger} into Eq.~\eqref{eq:num-h-free-tot-final-app} yields the Hamiltonian as a linear combination of Pauli strings. We do not show this substitution explicitly; it is carried out by the implementation described in \hyperref[sec:implementation]{App.~\ref*{sec:implementation}}.

\subsubsection{Discrete symmetries and obstacles in 2\,+\,1\textit{D}}\label{sec:d-symmetries}

In this section, we discuss the discrete symmetries of free Wilson fermions in $2+1D$ and their consequences for ADAPT-VQE. Charge conjugation will be used to construct a symmetry-preserving operator pool in \autoref{sec:introducting symmetries}. Moreover, we show that, although time reversal symmetry of the physical system is broken, the qubit Hamiltonian obtained from the JWT on a $2\times 2$ lattice is time-reversal symmetric in the sense of \autoref{sec:time-reversal-symmetry}, i.e., real valued, which leads to the obstacle discussed there.

The discrete symmetries of massive fermions are special in two spatial dimensions. Specifically, for $m\neq 0$ parity \parity\,as well as time reversal \timereversal\,are explicitly broken. 
The only unbroken discrete symmetry is charge conjugation \chargeconjugation, it acts on the two-component Spinor $\Psi$ as
\begin{equation}
    \chargeconjugation \Psi\chargeconjugation^{-1}=\sigma_x \left(\Psi^\dagger\right)^\intercal. \label{eq:cpt-c-psi}
\end{equation}
This can be seen by following a similar line of reasoning as \textcite[\sect{2.6, 4.2, 5.5}]{Weinberg:1995} for the here chosen gamma matrices in Eq.~\eqref{eq:num-2+1D-gamma-choice}.
From this, the transformation behavior of the fermionic modes $\Phi_j=\Phi(m_x,n_y)$ introduced in Eq.~\eqref{eq:num-psi-phi} follows to be:
\begin{equation*}
    \chargeconjugation \Phi(m_x,n_y)\chargeconjugation^{-1}=
    \begin{dcases}
        \Phi^\dagger(m_x+1,n_y) &m_x \text{ even},\\
        \Phi^\dagger(m_x-1,n_y) &m_x \text{ odd}.
    \end{dcases}
\end{equation*}
Under the ordering introduced in Eq.~\eqref{eq:qc-j-snake} this translates to
\begin{equation}
    \chargeconjugation \Phi_j\chargeconjugation^{-1}
    =
    \begin{dcases}
        \Phi^\dagger_{j+1} &j \text{ even},\\
        \Phi^\dagger_{j-1} &j \text{ odd}.
    \end{dcases}\label{eq:cpt-c-phi-j}
\end{equation}

Under charge conjugation, a general basis state $\ket{i_0,\dots, i_{N-1}}$ transforms blockwise on pairs of qubits $(k,k+1)$, with $k$ even. Within each block,
\begin{align}
    \begin{aligned}
    \ket{00} &\overset{\chargeconjugation}{\rightarrow} \ket{11}, & \ket{01} &\overset{\chargeconjugation}{\rightarrow} \ket{01}, \\
    \ket{10} &\overset{\chargeconjugation}{\rightarrow} -\ket{10}, & \ket{11} &\overset{\chargeconjugation}{\rightarrow} \ket{00}.
    \end{aligned}
    \label{eq:qc-c-action-on-block}
\end{align}
This transformation behavior follows from Eq.~\eqref{eq:cpt-c-phi-j} by identifying the states of minimal and maximal charge in configuration and momentum space of the lattice model.

As already mentioned time reversal symmetry of the physical system is broken here. Despite this, the resulting qubit Hamiltonian after the JWT can be time reversal symmetric. 
This is because time reversal symmetry on the level of the qubits does not involve spinor components. 
Therefore, the transformation is defined as a complex conjugation in the computational basis; this implies that the qubit Hamiltonian has real valued components in the computational basis.

More precisely, the JWT convention adopted in \hyperref[sec:jwt-a]{App.~\ref*{sec:jwt-a}} maps the fermionic modes $\Phi_j^{(\dagger)}$ to multi-qubit operators with real-valued components. Consequently, the only possible imaginary contribution to $H_{\text{Free}}$ in Eq.~\eqref{eq:num-h-free-tot-bc} comes from the term with the gamma matrices. For a $2\times 2$~lattice with periodic boundary conditions, this term cancels exactly, since neighboring lattice sites are effectively connected twice.

In summary, for a $2\times 2$~lattice of free Wilson fermions, time-reversal symmetry is broken in the physical system but present in the mapped Hamiltonian when viewed as a qubit Hamiltonian.

\subsection{Numerical setup and VQE baseline}\label{sec:numerics-setup}

In what follows we set the numerical setup for our variational simulations and discuss VQE as a baseline for accessing properties of ground and low excited states of the theory of free Wilson fermions.

We consider the total charge equal zero sector
\begin{equation}
    Q\ket{\Psi}=0,
\end{equation}
where the charge operator $Q$ is given by 
\begin{equation}
    Q=\sum\limits_{m_x=0}^{M_x-1} \sum\limits_{n_y=0}^{N_y-1}\brac*{\Phi^\dagger(m_x, n_y) \Phi(m_x, n_y) - \dfrac{1}{2}}, \label{eq:cpt-q-2+1d-disc}
\end{equation}
and the one acting on qubits
\begin{equation}
    Q= \sum\limits_{j=0}^{N-1}\dfrac{1}{2}\brac*{-\pz\qubit{j}}\label{eq:cpt-q-2+1d-disc-qubits}.
\end{equation}
We enforce this restriction by adding a penalty term of the form 
\begin{equation}
    H_{Q,\text{penalty}}=\lambda_Q Q^2,
\end{equation}
to the Hamiltonian $H_{\text{Free}}$.
With this the cost function for these routines is now given by
\begin{equation}
    f_c\brac*{\Vec{\theta}}=\left<O\right>_{\Vec{\theta}}=\bra{\psi\brac*{\theta}}\brac*{H_{\text{Free}}+\lambda_Q Q^2}\ket{\psi\brac*{\theta}}.
    \label{eq:cost-function-specific}
\end{equation}

The coefficient $\lambda_Q$ is chosen to be $\lambda_Q=100$ in the following 
and we fix the Wilson parameter to $r=1$ \footnote{We numerical verified that $\lambda_Q=100$ is sufficient to push states with finite charge outside the low energy spectrum.}.

We will here consider a lattice of size $N_x=N_y=2$ with periodic boundary conditions. With this the fermionic modes introduced in Eq.~\eqref{eq:num-psi-phi} are defined on a lattice of size $(M_x=4) \times (N_y=2)$.

The system is thus defined on $N_{\text{qubits}} = 8$ qubits. The resulting $2^8\times 2^8$ Hamiltonian matrix is small enough for exact diagonalization, 
and the energy eigenvalues obtained in this way serve as a reference for verifying VQE and ADAPT-VQE. All results for these hybrid quantum-classical algorithms were obtained by simulating an ideal, noise-free quantum computer with the \verb|StatevectorEstimator| provided by \verb|Qiskit|~\cite{qiskit:2024}.

\subsubsection{VQE baseline results}\label{sec:naive-greedy}

We first apply VQE, as introduced in \autoref{sec:vqe-alg}, to prepare the ground state, providing a baseline against which to compare ADAPT-VQE. This requires an initial state $\ket{\psi_0}$ and a parametrized circuit $U\brac{\vec{\theta}}$. As the initial state, we choose
\begin{equation}
    \ket{\psi_{0,0}}=\ket{01010101}, \label{eq:avqe-state-0-vqe}
\end{equation}
We build the parametrized circuit out of the layers previously used by \textcite{Schuster:2024-06} for the $1+1D$ case. Specifically, the ansatz is composed of gates generated by the Hermitian operators $\{\mathbf{XX+YY}, \mathbf{Z}\}$. Each layer consists of two successive sets of two-qubit gates of the form $\exp\left(-i\theta(\mathbf{XX+YY})/2\right)$, applied first to non-overlapping pairs of qubits starting at even indices and then to non-overlapping pairs starting at odd indices. This is followed by single-qubit $\mathbf{Z}$-rotations applied to each qubit. Each of these gates has its own parameter. The initial state is chosen in the charge-zero sector and the operators commute with the charge operator. The initial parameters $\theta_i^{(0)}$ are sampled uniformly in~$(0, 2\pi)$~$\forall i$.

\begin{figure}[htbp]
    \centering
    \includegraphics[width=\linewidth]{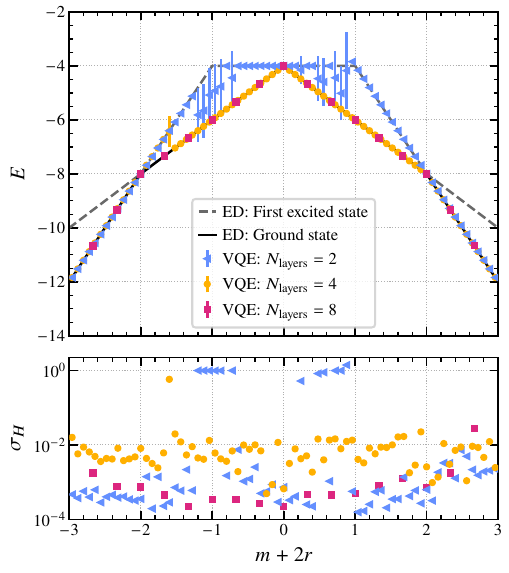}
    \caption[Energy against shifted mass, VQE for varying number of circuit layers]{Energy~(top) and standard deviation of the Hamiltonian on a logarithmic scale~(bottom) against shifted mass $m+2r=m+2$. We show the ground (solid black line) and first excited states (dashed gray line) obtained from ED as well as the final states obtained from VQE for a parametrized circuit consisting of $N_{\text{layers}}=2$, $4$ and $8$ of the layers, corresponding to blue triangles, orange dots and magenta squares respectively.
    The gradient-free optimizer used is COBYLA with an iteration limit of $I_{\text{max}}=\num{20000}$ and a target accuracy of $\varepsilon_{\text{opt}}=\num{1e-4}$.
    }
    \label{fig:vqe_varying_layer_count}
\end{figure}

We can now perform the VQE optimization for varying values of the mass parameter $m$.
The final energy values obtained in this way are shown in \autoref{fig:vqe_varying_layer_count} for varying numbers of layers in the ansatz of $N_{\text{layers}}=2$, $4$ and $8$.
In addition, we also show the associated final standard deviation of the Hamiltonian with respect to the final state
\[\sigma_H = \sqrt{\operatorname{Var}\brac*{H_{\text{Free}}}}=\sqrt{\left<H_{\text{Free}}^2\right>-\left<H_{\text{Free}}\right>^2}, \] in the bottom panel, and as the error-bars in the top panel. It should be stressed that the latter are not to be interpreted as statistical errors: a smaller value of this quantity corresponds to a higher fidelity in the approximation of an energy eigenstate.

We observe that a parametrized circuit $U(\vec{\theta})$ consisting of $N_{\text{layers}}=8$ layers reconstructs the ground state in all parameter regions. For $N_{\text{layers}}=4$ layers we see a reconstruction of the ground state in all parameter regions except for $-2<m+2r<-1$.

Since the initialization in step~\ref{enum:vqe-init-params} of VQE is performed randomly, our results might depend on this initialization. By performing multiple optimization runs with varying random seeds we verified this. We observe a dependence in the ground state reconstruction on the parameter initialization for $-2<m+2r<-1$ and $1<m+2r<2$. By performing multiple (\order{5}) optimization runs and only using the smallest final energy value for each value of the mass parameter, we can thus reconstruct the ground state using $N_{\text{layers}}=4$ layers in the parametrized circuit $U(\vec{\theta})$.

With only $N_{\text{layers}}=2$ layers, the ground state is successfully reconstructed only for $m+2r<-2$ and $m+2r>2$. This is the first instance in which the level crossings at $m+2r=-2,0,+2$ cause the first excited state to be predominantly reconstructed in the intermediate region $-2<m+2r<2$, a pattern that recurs in the following. The integer values of the shifted mass at which these level crossings occur agree with the theoretical prediction of \textcite{Golterman:1992-08}.

Furthermore, from the bottom panel, we see that in this region multiple optimization runs do not result in an energy eigenstate. For these the standard deviation is $\order{1}$, and the obtained energy values do not match a specific reference energy level from ED in the top panel.

\subsection{ADAPT-VQE results}

In this section we discuss our analysis of ADAPT-VQE applied to $2+1D$ Wilson fermions. After pointing out fundamental limitations of ADAPT-VQE in reproducing the same circuit of VQE, we analyze different operator pools and the role of symmetries in our setup.

\subsubsection{Failure for real valued operators}\label{sec:fail-real-op}

As a first test, we check whether ADAPT-VQE can reconstruct the parametrised circuit used for VQE. To this end, we adopt a so-called \enquote{greedy} approach and provide only the elementary components, namely single- and two-qubit operators. The initial state is again $\ket{\psi_{0,0}}=\ket{01010101}$, and the operator pool $P^{(0)}$ is given by the set
\begin{align}
    P^{(0)}_{1,i}&\coloneqq\set{-\frac{1}{2}\pz\qubit{i}}\nonumber,\\
    P^{(0)}_{2,i}&\coloneqq
    \set{-\frac{1}{2}\brac*{\px\qubit{i-1}\px\qubit{i}+\py\qubit{i-1}\py\qubit{i}}}\nonumber,\\
    P^{(0)}&\coloneqq \brac*{\bigcup \limits_{i=0}^{N_{\text{qubits}}-1}P^{(0)}_{1,i}} \cup \brac*{ \bigcup \limits_{i=1}^{N_{\text{qubits}}-1}P^{(0)}_{2,i}},
\end{align}
such that once exponentiated these operators result in the rotation gates $R_{\px\px+\py\py}$ and $R_{\pz}$.

Given the initial state $\ket{\psi_{0,0}}$ and operator pool~$P^{(0)}$, it is now clear that the optimization routine will immediately exit at step~\ref{enum:avqe-exit-1} without adding any gate to the ansatz. This follows from the fact that the qubit Hamiltonian is time-reversal symmetric, the initial wave function is real valued, and all the operators in the pool are real: all the gradients~\eqref{eq:num-avqe-grad} are zero at the first iteration.

We then observe that ADAPT-VQE is not able to reconstruct arbitrary circuits. In particular, circuits that are known to be able to reach the ground state from VQE simulations are generally beyond the reach of ADAPT-VQE.

There are multiple ways of avoiding this problem. One option is to choose a non-real-valued initial wavefunction~\footnote{Even though this is not sufficient on its own, as discussed in the third statement of~\autoref{sec:time-reversal-symmetry}.}.
A second option is to add complex-valued operators to the operator pool. In the next sections we will explore the latter option in detail. Also, since the explicit components of the Hamiltonian, once mapped to qubits, depend on the conventional phase choice in the JWT, another option is to choose a different local phase convention in the JWT.

\subsubsection{Minimal pool}\label{sec:greedy-pool}

So far, the ansatz was designed to start in and remain within the sector with vanishing total charge. We now relax this constraint, since no Hermitian single-qubit operator with non-real components commutes with the charge operator.

Specifically, we chose the initial state to be the equal weight superposition state
\begin{equation}
    \ket{\psi_{0,1}}=\frac{1}{\sqrt{2^8}}\sum\limits_{k=0}^{2^8-1}\ket{k}, \label{eq:avqe-state-1}
\end{equation}
which can easily be prepared on a quantum computer by applying Hadamard gates to all qubits.
The operator pool we will use now is given by
\begin{align}
    P^{(1)}_{1,i}&\coloneqq\set{-\frac{1}{2}\py\qubit{i}}\nonumber,\\
    P^{(1)}_{2,i}&\coloneqq
    \bigg\{-\frac{1}{2}\brac*{\px\qubit{i-1}\py\qubit{i}-\py\qubit{i-1}\px\qubit{i}}\bigg\}\nonumber,\\
    P^{(1)}&\coloneqq \brac*{\bigcup \limits_{i=0}^{N_{\text{qubits}}-1}P^{(1)}_{1,i}} \cup \brac*{ \bigcup \limits_{i=1}^{N_{\text{qubits}}-1}P^{(1)}_{2,i}}.\label{eq:avqe-pool-2}
\end{align}

This is chosen to contain operators that are not real valued. The gates $R_{\px\py-\py\px}$ resulting from $P^{(1)}_{2,i}$ have been implemented using a circuit from \textcite{Farrell:2024-04}.
One can show that these two-qubit operators commute with the charge operator, while the single qubit one does not commute with $Q$.

\begin{figure}[htbp]
    \centering
    \includegraphics[width=\linewidth]{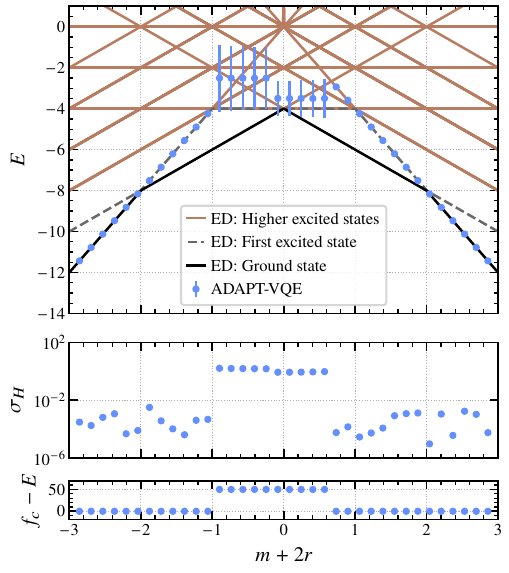}
    \caption[Energy against shifted mass, ADAPT-VQE using operator pool~$P^{(1)}$ and initial state~$\ket{\psi_{0,1}}$]{Energy~(top), standard deviation of the Hamiltonian~(middle), and penalty term~(bottom) against shifted mass $m+2r=m+2$. We show the ground, first, and lowest lying higher excited states obtained from ED, as well as the final states obtained from ADAPT-VQE on a $2\times 2$~lattice.
    The threshold parameters of ADAPT-VQE are set to $\varepsilon=\num{1e-3}$ and $\varepsilon_\theta=\num{1e-3}$. The VQE subroutine uses SLSQP as the optimizer with an iteration limit of $I_{\text{max}}=\num{20000}$ for each ADAPT-VQE iteration and a target accuracy of $\varepsilon_{\text{opt}}=\num{1e-6}$.
    }
    \label{fig:adapt_vqe_3_e_m}
\end{figure}

We can now perform the ADAPT-VQE optimization with this choice of operator pool $P^{(1)}$
and initial state $\ket{\psi_{0,1}}$ for varying values of the mass parameter $m$.
The final energy values and standard deviations of the Hamiltonian obtained in this way are shown in \autoref{fig:adapt_vqe_3_e_m}. In addition, we also show the penalty term $\lambda_Q \left<Q^2\right>\overset{\ref{eq:cost-function-specific}}{=}f_c-E$ in the bottom panel.

First, we note that with the operator pool $P^{(1)}$ and initial state $\ket{\psi_{0,1}}$, ADAPT-VQE reproduces the ground state for $m+2r<-2$ and $m+2r>2$. Towards the intermediate region, we encounter the second instance in which the level crossings discussed in \autoref{sec:naive-greedy} cause the first excited state to be reconstructed, instead of the ground state. This illustrates the second limitation of ADAPT-VQE described in \autoref{sec:higher-excited-states}. Where either the ground or the first excited state is reconstructed, the fidelity is slightly improved compared with the VQE baseline of \autoref{sec:naive-greedy}: the final standard deviation of the Hamiltonian in the middle panel is $\order{10^{-3}}$.

In the remaining region $-1<m+2r<1$, the ten optimization runs fail to reconstruct the ground state. The middle panel shows that these runs do not converge to an energy eigenstate, as the standard deviation is $\order{1}$ and the resulting energies do not match any reference level from ED in the top panel. Finally, the bottom panel shows non-vanishing final penalty terms for these runs, indicating that the final states lie outside the zero-charge sector. This is not surprising, since the ansatz is not constrained to this symmetry sector.

For completeness, we briefly outline different strategies we pursued to improve the convergence to the ground state while retaining a minimal pool. None of these led to a significantly better convergence, as it is instead the case of incorporating charge conjugation symmetry into the ansatz, as discussed in detail in the next section.

Adding Gaussian noise to the gradient measurements in step~\ref{enum:avqe-grad}, intended to escape vanishing gradients at excited eigenstates, gave at most a marginal improvement for narrow noise distributions. Initial states $\ket{\psi_0}$ with wave functions that are not real up to a global phase, chosen to circumvent the problem of \autoref{sec:time-reversal-symmetry}, consistently worsened the results. Among the optimizers COBYLA, COBYQA, and SLSQP, the quadratic ones increased the number of runs reaching the ground state, with SLSQP converging fastest; loosening the target accuracy $\varepsilon_{\text{opt}}$ beyond its default value of $\num{1e-6}$ favored higher excited states, whereas tightening it gave no benefit. Lowering the thresholds $\varepsilon$ and $\varepsilon_\theta$ below $\num{1e-3}$ had no noticeable effect. Computing the gradients in step~\ref{enum:avqe-grad} from $H_{\text{Free}}$ without the penalty term, so that the operator selection is not dominated by the penalty, caused more runs to end in higher excited states or outside the zero-charge sector. Finally, we restricted the pool to charge-preserving operators. Starting in the zero-charge sector with only charge-preserving two-qubit operators produced states matching a specific energy level without being energy eigenstates, whereas restricting only the two-qubit operators while retaining non-charge-preserving single-qubit operators performed best. The results shown above were obtained with this pool and the SLSQP optimizer.

\subsubsection{
Incorporating additional symmetries into the operator pool}\label{sec:introducting symmetries}

In the previous sections, we followed a greedy approach in which only operators acting on one and two qubits were considered. Moreover, the only systematic constraint on the construction of the operator pool so far has been the limitation of ADAPT-VQE discussed in~\autoref{sec:time-reversal-symmetry}. In the following, we incorporate charge conjugation symmetry $\chargeconjugation$, discussed in~\autoref{sec:d-symmetries}, into the operator pool using the scheme presented in~\autoref{sec:symmetries}.

Since charge conjugation is an unbroken symmetry here, as discussed in \autoref{sec:d-symmetries}, the ground state is a $\chargeconjugation$ eigenstate. The constructed operator pool can therefore, in principle, reproduce the ground state, provided the initial state has the same $\chargeconjugation$ eigenvalue as the target.

An approach to construct hermitian operators satisfying Eq.~\eqref{eq:comm-general-symmetry} is using invariant parts of the Hamiltonian $H_{\text{Free}}$ given in Eq.~\eqref{eq:num-h-free-tot-bc}, similar to what has been done in~\textcite{Farrell:2024-04}. Specifically by introducing
\begin{align*}
    O_1(n)&\coloneqq \Psibar(n) \Psi(n) = O_1^\dagger(n),\\
    O_2(n,j)&\coloneqq \Psibar(n) \iu \gamma^j\Psi\brac*{n+\widehat{j}},\\
    O_3(n,j)&\coloneqq \Psibar(n) (-\iu) \gamma^j\Psi\brac*{n-\widehat{j}}=O_2^\dagger\brac*{n-\widehat{j},j},\\
    O_4(n,j)&\coloneqq \Psibar(n) \Psi\brac*{n+\widehat{j}},\\
    O_5(n,j)&\coloneqq \Psibar(n) \Psi\brac*{n-\widehat{j}}=O_4^\dagger\brac*{n-\widehat{j},j},
\end{align*}
we see that 
\begin{align*}
    O_1(n),\qquad &O_2(n,j)+O_2^\dagger(n,j)\quad \\
    \text{and} \quad &O_4(n,j)+O_4^\dagger(n,j)
\end{align*}
commute with charge conjugation. Once we express these operators in terms of the fermionic modes and perform the JWT, we see that most of the former operators are real valued when acting on qubits. This would therefore lead to the limitation discussed in~\autoref{sec:time-reversal-symmetry}. To circumvent this problem we can build the following operators
\begin{align*}
    A_1(n)&=O_2(n,1)+O_2^\dagger(n,1),\\
    A_2(n)&=\iu\comm{O_1(n)}{O_2(n,2)+O_2^\dagger(n,2)},\\
    A_3(n)&=\iu\comm{O_1(n)}{O_4(n,1)+O_4^\dagger(n,1)},\\
    A_4(n)&=\iu\comm{O_1(n)}{O_4(n,2)+O_4^\dagger(n,2)},
\end{align*}
which are guaranteed to be imaginary valued and commute with charge conjugation~\footnote{Another option would be to follow the approach of rephasing the JWT as mentioned in \autoref{sec:fail-real-op}}.

The constructed operators commute with $Q$. The charge conjugation and charge invariant operator pool $P^{(2)}$ is now given by
\begin{equation}
    P^{(2)}=\set{A_i(n)\middle| i \in \set{1,2,3,4}, n \in \Lambda}.\label{eq:avqe-pool-3}
\end{equation}
The corresponding multi qubit gates 
\begin{align}
U_j^{\left[n_x, n_y\right]}(\theta)=\euler^{\iu A_j(n_x, n_y)}
\end{align}
can be decomposed into two qubit gates by following the same decomposition and compression as introduced by~\textcite{Algaba:2024-04}.

The only remaining task now is to choose the initial state $\ket{\psi_0}$ to have total charge zero and to be an eigenstate of charge conjugation with the same eigenvalue as the ground state~\footnote{Due to the phase convention adopted in \autoref{sec:d-symmetries} this eigenvalue is \emph{not} fixed to $+1$ but it can take on the values~$\alpha_{C,\Omega}=\pm 1$.}\!\!.
From the transformations given in \autoref{sec:d-symmetries} we see that the following states are valid choices: 
\begin{align}
    &\ket{\psi_{0,0}}=\ket{01010101}  \qquad \chargeconjugation \ket{\psi_{0,0}} \overset{\ref{eq:qc-c-action-on-block}}{=} +1 \ket{\psi_{0,0}},\label{eq:avqe-state-0}\\
    &\ket{\psi_{0,2}}=\ket{01010110} \qquad \chargeconjugation \ket{\psi_{0,2}} \overset{\ref{eq:qc-c-action-on-block}}{=} -1 \ket{\psi_{0,2}},\label{eq:avqe-state-2}\\
    &\ket{\psi_{0,3}}=\ket{10010101} \qquad \chargeconjugation \ket{\psi_{0,3}} \overset{\ref{eq:qc-c-action-on-block}}{=} -1 \ket{\psi_{0,3}}.\label{eq:avqe-state-3}
\end{align}
These states are all in the $Q=0$ sector.

\subsubsection{Numerical results using symmetry-adapted operator pools}

In what follows we compare the results obtained from the operator pool~$P^{(1)}$ in the previous sections, here named ADAPT-VQE 1, where charge conjugation is not imposed as a symmetry, with the new runs, namely ADAPT-VQE 2, 3 and 4, obtained using the operator pool~$P^{(2)}$ and the initial states $\ket{\psi_{0,0}}$, $\ket{\psi_{0,2}}$ and $\ket{\psi_{0,3}}$, respectively.

\begin{figure}[htbp]
    \centering
    \includegraphics[width=\linewidth]{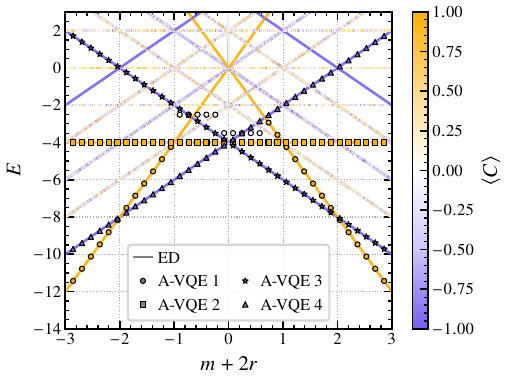}
    \caption[Energy and charge conjugation against shifted mass, ADAPT-VQE using operator pool~$P^{(1)}$ and initial state~$\ket{\psi_{0,1}}$, operator pool~$P^{(2)}$ and initial state~$\ket{\psi_{0,0}}$]{Energy against shifted mass, where the colors of the lines and points indicate the corresponding charge conjugation expectation values. We show the ground, first, and lowest lying higher excited states obtained from ED, as well as the final states obtained from ADAPT-VQE on a $2\times 2$~lattice.
    The threshold parameters of ADAPT-VQE are set to $\varepsilon=\num{1e-3}$ and $\varepsilon_\theta=\num{1e-3}$. The VQE subroutine uses SLSQP as the optimizer with an iteration limit of $I_{\text{max}}=\num{20000}$ for each ADAPT-VQE iteration and a target accuracy of $\varepsilon_{\text{opt}}=\num{1e-6}$.
    }
    \label{fig:adapt_vqe_3-4_c_e_m}
\end{figure}

Specifically, the final energy values and corresponding charge conjugation expectation values~$\left<\chargeconjugation\right>$ are shown in~\autoref{fig:adapt_vqe_3-4_c_e_m}.

First we note that the variance of $\chargeconjugation$ is not shown since it is redundant in the sense that
\begin{equation*}
    \operatorname{Var}\brac*{\chargeconjugation}=\bra{\psi} \underbrace{\chargeconjugation^2}_{=\identity} \ket{\psi} - \brac*{\bra{\psi}\chargeconjugation\ket{\psi}}^2 =  1 - \left<\chargeconjugation\right>^2.
\end{equation*}
Thus, a charge conjugation expectation value near $\pm 1$ indicates that the final state is close to a charge conjugation eigenstate.
Moreover, ADAPT-VQE~2, 3, and~4 reach a final standard deviation of the Hamiltonian of $\sigma_H=\order{10^{-4}}$ over the entire mass range considered, i.e., they always converge to an energy eigenstate, even where this is an excited state. For ADAPT-VQE~1, the standard deviation is shown in \autoref{fig:adapt_vqe_3_e_m}.

We begin by analyzing the reference ED results. As expected, the ground state is an eigenstate of charge conjugation. 
Note that we fix the phase of $\chargeconjugation$ through its action on the reference state, Eq.~\eqref{eq:qc-jwt-reference-state}, rather than requiring the ground state to have eigenvalue $+1$, as is commonly done. With this convention, each eigenvalue is tied to a specific state, such that a change of the ground-state eigenvalue directly signals a level crossing between states with different $\chargeconjugation$ eigenvalues. Indeed, the ground-state eigenvalue changes at $m+2r=-2$ and $m+2r=2$. Consequently, close to these crossings, the first excited state is the continuation of the ground state on the other side of the crossing and is thus also a $\chargeconjugation$ eigenstate, which is the case for $m+2r<-1$ and $m+2r>1$.
Some of the higher excited states are $\chargeconjugation$ eigenstates as well, although most energy levels do not exhibit a definite $\chargeconjugation$ eigenvalue over an extended parameter region. This is because these levels are degenerate and the diagonalization involves only the Hamiltonian, so the resulting state may either be a $\chargeconjugation$ eigenstate or an arbitrary superposition of $\chargeconjugation$ eigenstates with different eigenvalues.

ADAPT-VQE 2 is the only run forced to the $\langle\chargeconjugation\rangle=+1$ sector. As evident from the plot, it never reconstructs the ground state, but instead consistently converges to an excited state. In the parameter region $-1<m+2r<+1$ this corresponds to the first excited state, while for larger and smaller values of the shifted mass this is a higher excited state. This run can be seen as an example of what was discussed in~\autoref{sec:higher-excited-states}, namely a premature exit when an excited state is reached~\footnote{Furthermore, it is possible that the expressivity of the operator pool $P^{(2)}$ is not sufficient to capture the lower lying energy eigenstates with the same charge conjugation eigenvalue as this operator pool does not include one, two or three qubit operators and corresponding gates.}.

ADAPT-VQE~3 and~4 yield similar but not equivalent results. Both runs are restricted to the $\langle\chargeconjugation\rangle=-1$ sector and reconstruct energy levels energy levels that are mirror-symmetric under $m+2r\to -(m+2r)$. ADAPT-VQE~3 reproduces the ground state for $0<m+2r<2$ and the first excited state for larger values of $m+2r$, whereas ADAPT-VQE~4 reproduces the ground state for $-2<m+2r<0$ and the first excited state for smaller values. Consequently, wherever one of the two runs converges to a higher excited state, the other reconstructs either the ground state or the first excited state.

We observe that no combination of the considered operator pools and initial states can reproduce the ground state for \emph{all} values of the mass parameter $m$. However, by combining the different results in~\autoref{fig:adapt_vqe_3-4_c_e_m}, we can observe that for all mass parameter values these combined results can reconstruct the ground \emph{and} first excited states. 

Specifically, the ground state is reconstructed for $m+2r<-2$ and $m+2r>2$ using the operator pool~$P^{(1)}$ of Eq.~\eqref{eq:avqe-pool-2} with the initial state~$\ket{\psi_{0,1}}$ of Eq.~\eqref{eq:avqe-state-1}, and for $-2<m+2r<2$ using the operator pool~$P^{(2)}$ of Eq.~\eqref{eq:avqe-pool-3} with the initial states~$\ket{\psi_{0,2}}$ and~$\ket{\psi_{0,3}}$ of Eqs.~\eqref{eq:avqe-state-2} and~\eqref{eq:avqe-state-3}, respectively.

\section{Conclusions}
\label{sec:conclusions}

In this work, we investigated whether the success of ADAPT-VQE in preparing ground states of $1+1D$ lattice field theories can be carried over to higher dimensions. To this end, we identified two general constraints of the algorithm. First, if the qubit Hamiltonian is time-reversal symmetric, i.e., real valued in the computational basis, and the initial state is real, the gradients of all real-valued pool operators vanish identically. Such operators are never selected, and ADAPT-VQE terminates immediately if the pool contains only such operators. This qubit-level symmetry need not coincide with time-reversal symmetry of the physical system, since the latter also acts on the spinor components of the fields. Second, every energy eigenstate is a stationary point of the energy, so ADAPT-VQE can converge prematurely once an excited state has been prepared. We further discussed how incorporating symmetries into the operator pool and the initial state affects the convergence of the algorithm.

Revisiting examples from the literature, we found that the successful application of SC-ADAPT-VQE to the Schwinger model~\cite{Farrell:2024-04} relies on imposing qubit-level time-reversal symmetry on the operator pool, which, in light of our analysis, is a necessity rather than an option for ADAPT-VQE to operate at all. The same obstacle also arises outside of lattice field theory, as we showed for an operator pool proposed for supersymmetric quantum mechanics~\cite{Kerfoot:2025mvx}.

We then presented a first application of ADAPT-VQE to a lattice field theory in $2+1D$, namely free Wilson fermions on a $2\times 2$ lattice. Although physical time-reversal symmetry is broken for massive fermions in $2+1D$, the qubit Hamiltonian obtained from the JWT on this lattice is real valued. Consequently, ADAPT-VQE cannot reconstruct a circuit that VQE uses successfully to reach the ground state, since all of its operators are real valued. With a minimal pool of complex-valued one- and two-qubit operators, ADAPT-VQE reconstructs the ground state far from the level crossings, but converges to the first excited state or fails to reach an eigenstate close to them. Incorporating charge conjugation into the operator pool improves the reconstruction of energy eigenstates, but no single combination of operator pool and initial state reproduces the ground state for all values of the mass. By combining the different runs, however, both the ground and the first excited state can be reconstructed over the full mass range considered.

These findings suggest some practical guidelines for applying ADAPT-VQE to lattice field theories beyond $1+1D$. Before choosing an operator pool, one should check whether the qubit Hamiltonian is real valued, which depends not only on the theory but also on the lattice size and the phase convention of the fermion-to-qubit mapping. If it is, the pool has to contain operators with non-vanishing imaginary parts, ideally purely imaginary ones, such that the wave function remains real. Since convergence of ADAPT-VQE does not guarantee that the ground state has been reached, the final state should be checked, for example, via the variance of the Hamiltonian and the expectation values of the relevant symmetries. Finally, in theories with level crossings, the symmetry sector of the ground state can change as a function of the parameters, so different symmetry sectors should be targeted by suitable initial states and the results compared.

The problem of premature convergence at excited states may be mitigated by replacing the local, gradient-based selection criterion of ADAPT-VQE with a non-local one. An example is the greedy gradient-free adaptive VQE (GGA-VQE)~\cite{Feniou:2023gga}, which, for each pool operator, determines the energy as a one-dimensional function of the corresponding parameter and selects the operator and angle that yield the lowest energy. Since this criterion does not rely on the gradient at the current state, it can in principle leave a stationary point such as an excited eigenstate, provided the pool contains an operator along which the energy decreases at finite angle. This makes non-local selection criteria a promising direction for applying adaptive variational algorithms to lattice field theories with level crossings.

Several directions remain for future work. On larger lattices, the gamma-matrix contribution to the qubit Hamiltonian no longer cancels, so the time-reversal properties of the qubit Hamiltonian, and thus the requirements on the operator pool, change. Moreover, incorporating charge conjugation into the pool directly leads to operators acting on more qubits, and disentangling the effects of symmetries and of larger operators is left for future studies. Finally, it will be important to extend our analysis to interacting theories, in particular gauge theories in $2+1D$ and $3+1D$, and to study the performance of these algorithms on noisy quantum hardware.

\begin{acknowledgments}
\noindent
The authors acknowledge insightful discussions with Emanuele Mendicelli, Emil Otis Rosanowski and Simran Singh. This project was supported by the Deutsche Forschungsgemeinschaft (DFG, German Research Foundation) as part of the CRC 1639 NuMeriQS -- project no.\ 511713970 and under Germany's Excellence Strategy -- Cluster of Excellence ``Color meets Flavor'' (CmF) EXC 3107 -- 533766364 and Cluster of Excellence ``Matter and Light for Quantum Computing'' (ML4Q) EXC 2004/2 -- 390534769. This work is part of the Quantum Computing for High-Energy Physics (QC4HEP) working group. The authors gratefully acknowledge the granted access to the Marvin cluster hosted by the University of Bonn.

This work is supported with funds from the Ministry of Science, Research and Culture of the State of Brandenburg within the Center for Quantum Technology and Applications (CQTA). 
\begin{center}
    \includegraphics[width = 0.08\textwidth]{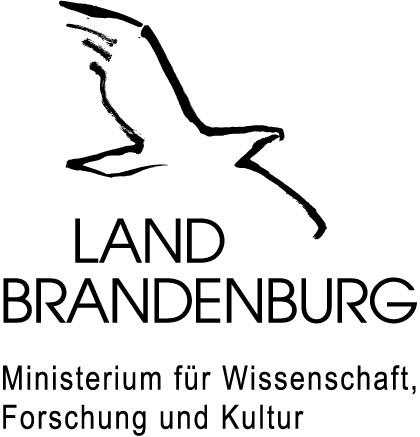}
\end{center}

We used AI assistance (Claude Opus 5.5) for language editing and to assist with plotting scripts. All output was checked by the authors, who take full responsibility for the content of this work.

\end{acknowledgments}

\appendix

\section{Free Wilson Hamiltonian in terms of fermionic modes} \label{sec:ham-ferm-modes}

In order to express the Hamiltonian in Eq.~\eqref{eq:num-h-free-tot-bc} in terms of the fermionic modes $\Phi(m_x,n_y)$, we choose the following gamma matrices:
\begin{align}
    \gamma^0 &= \sigma_z = \brac*{\gamma^0}^\intercal=\brac*{\gamma^0}^*,\nonumber\\
    \gamma^1&=-\iu\sigma_y= -\brac*{\gamma^1}^\intercal=\brac*{\gamma^1}^*,\label{eq:num-2+1D-gamma-choice}\\
    \gamma^2&=-\iu\sigma_x = \brac*{\gamma^2}^\intercal=-\brac*{\gamma^2}^*.\nonumber
\end{align}
Here $\sigma_x$, $\sigma_y$, and $\sigma_z$ denote the Pauli matrices.

Inserting these into Eq.~\eqref{eq:num-h-free-tot-bc} one obtains
\begin{widetext}
    \begin{align}
        \begin{aligned}
            H_{\text{Free}}\label{eq:num-h-free-tot-final-app}
            &=\sum\limits_{n_x=0}^{N_x-1}\sum\limits_{n_y=0}^{N_y-1}\Bigg[-\dfrac{1}{2}\bigg\{\\
            &\phantom{=} r\big(B_1(2n_x,n_y)+B_1(2n_x+1,n_y)+B_2(2n_x,n_y)+B_2(2n_x+1,n_y)\big)\\
            &\phantom{=}+C_1(2n_x+1,n_y)+C_1(2n_x,n_y)+C_2(2n_x+1,n_y)+C_2(2n_x,n_y)+\hermitc\bigg\}\\
            &\phantom{=}+\left(m+2 r\right)(A(2n_x,n_y)+A(2n_x+1,n_y))\Bigg]\\
            &=\sum\limits_{m_x=0}^{M_x-1}\sum\limits_{n_y=0}^{N_y-1}\Bigg[
            \left(m+2 r\right)A(m_x,n_y)\\
            &\phantom{=}-\dfrac{1}{2}\bigg\{r\big(B_1(m_x,n_y)+B_2(m_x,n_y)\big)+C_1(m_x,n_y)+C_2(m_x,n_y)+\hermitc\bigg\}\Bigg],
        \end{aligned}
    \end{align}
\end{widetext}
where we have defined the abbreviations
\begin{align}
    &A(m_x,n_y)\label{eq:num-a-abbrev}\\&\coloneqq(-1)^{m_x}\Phi^\dagger(m_x,n_y)\Phi(m_x,n_y),\nonumber\\
    &B_1(m_x,n_y)\label{eq:num-b1-abbrev}\\&\coloneqq(-1)^{m_x}\Phi^\dagger(m_x,n_y)\Phi(m_x+2,n_y),\nonumber\\
    &B_2(m_x,n_y)\label{eq:num-b2-abbrev}\\&\coloneqq(-1)^{m_x}\Phi^\dagger(m_x,n_y)\Phi(m_x,n_y+1),\nonumber\\
    &C_1(m_x,n_y)\label{eq:num-c1-abbrev}\\&\coloneqq-\iu\bigg[\dfrac{1}{2}\left(1-(-1)^{m_x}\right)\Phi^\dagger(m_x,n_y)\Phi(m_x+1,n_y)\nonumber\\
    &\phantom{=}+\dfrac{1}{2}\left(1+(-1)^{m_x}\right)\Phi^\dagger(m_x,n_y)\Phi(m_x+3,n_y)\bigg],\nonumber\\
    &C_2(m_x,n_y)\label{eq:num-c2-abbrev}\\&\coloneqq-\dfrac{1}{2}\left(1-(-1)^{m_x}\right)\Phi^\dagger(m_x,n_y)\Phi(m_x-1,n_y+1)\nonumber\\
    &\phantom{=}+\dfrac{1}{2}\left(1+(-1)^{m_x}\right)\Phi^\dagger(m_x,n_y)\Phi(m_x+1,n_y+1). \nonumber
\end{align}

\section{Jordan--Wigner transformation}\label{sec:jwt-a}
{
\newcommand*\fermionop{\Phi}
\newcommand*\fermionvac{\omega}
To translate the fermionic degrees of freedom to qubits, we use a JWT.
The following derivation of the necessary properties of the JWT is based on Ref.~\cite{Verstraete:2005-09}.
A state of $N$ fermionic modes can be described as 
\begin{align*}
    \ket{\psi} &= \sum\limits_{i_1,\dots, i_N=0}^1 C_{i_1,\dots, i_N}\left(\fermionop_1^\dagger\right)^{i_1} \dots \left(\fermionop_N^\dagger\right)^{i_N} \ket{\fermionvac},
\end{align*}
with the anticommutation relations
\begin{equation*}
    \acomm{\fermionop_i}{\fermionop_j^\dagger}=\delta_{ij}\qquad \acomm{\fermionop_i}{\fermionop_j}=\acomm{\fermionop_i^\dagger}{ \fermionop_j^\dagger}=0,
\end{equation*}
and the reference state $\ket{\fermionvac}=\ket{0,\dots,0}$ is defined to satisfy:
\begin{equation*}
    \forall j \in \set{1,\dots, N}:\quad \fermionop_j\ket{\fermionvac}=0.
\end{equation*}
This fixes a specific ordering convention, such that by using the identification
\begin{equation*}
    \ket{i_1,\dots, i_N} \coloneqq \left(\fermionop_1^\dagger\right)^{i_1} \dots \left(\fermionop_N^\dagger\right)^{i_N} \ket{\fermionvac},
\end{equation*}
we can switch to using $N$ qubits to describe a specific state. Note that this involves a \emph{phase choice for every fermionic mode}.
Let us now derive the representation of the annihilation operator $\fermionop_k$:
\begin{widetext}
   \begin{align}
    \begin{aligned}
        \bra{j_1,\dots,j_N}\fermionop_k\ket{i_1,\dots, i_N}
        &=\bra{\fermionvac}\fermionop_N^{j_N}\dots \fermionop_1^{j_1}
        \fermionop_k
        \left(\fermionop_1^\dagger\right)^{i_1} \dots \left(\fermionop_N^\dagger\right)^{i_N} \ket{\fermionvac}\\
        &=\bra{\fermionvac}\fermionop_N^{j_N}\dots \fermionop_1^{j_1} 
        \delta_{i_k,1}\left(\prod\limits_{n<k}(-1)^{i_n}\right)
        \left(\fermionop_1^\dagger\right)^{i_1} \dots \left(\fermionop_k^\dagger\right)^{0}\dots\left(\fermionop_N^\dagger\right)^{i_N} \ket{\fermionvac}\\
        &=\delta_{i_k,1}\left(\prod\limits_{n<k}(-1)^{i_n}\right) \delta_{j_k,0}\cdot \underbrace{\bra{\fermionvac}\fermionop_N^{j_N}\dots \fermionop_k^{0}\dots \fermionop_1^{j_1}
        \left(\fermionop_1^\dagger\right)^{i_1} \dots \left(\fermionop_k^\dagger\right)^{0}\dots\left(\fermionop_N^\dagger\right)^{i_N} \ket{\fermionvac}
        }_{\prod\limits_{n\neq k} \delta_{j_n, i_n} }\\
        &=\delta_{i_k,1}\left(\prod\limits_{n<k}(-1)^{i_n}\right) \delta_{j_k,0} \prod\limits_{n\neq k} \delta_{j_n, i_n} \\
        &= \left(\prod\limits_{n<k} \pz^{(n)}_{j_n,i_n}\right) 
        \begin{pmatrix}
    0 & 1 \\
    0 & 0
    \end{pmatrix}_{j_k,i_k}^{(k)}
        \left(\prod\limits_{n>k} \identity^{(n)}_{j_n,i_n}\right)\\
        &= \left(\prod\limits_{n<k} \pz^{(n)}_{j_n,i_n}\right)
        \left(\dfrac{1}{2}(\px+\iu\py)\right)_{j_k,i_k}^{(k)} 
        \left(\prod\limits_{n>k} \identity^{(n)}_{j_n,i_n}\right).
        \end{aligned}
    \end{align} 
\end{widetext}

In summary we identified:
\begin{align*}
    \ket{\psi} &= \sum\limits_{i_1,\dots, i_N=0}^1 C_{i_1,\dots, i_N} \ket{i_1,\dots, i_N},\\
    \fermionop_k &= \left(\bigotimes\limits_{n<k} \pz^{(n)}\right) \otimes \left(\dfrac{1}{2}(\px+\iu\py)\right)^{(k)}\otimes \left(\bigotimes\limits_{n>k} \identity^{(n)}\right),\\
    \fermionop_k^\dagger &= \left(\bigotimes\limits_{n<k} \pz^{(n)}\right) \otimes \left(\dfrac{1}{2}(\px-\iu\py)\right)^{(k)}\otimes \left(\bigotimes\limits_{n>k} \identity^{(n)}\right).
\end{align*}

}

\section{Numerical implementation} \label{sec:implementation}
The code, which is available at \cite{nico-mt}, has been implemented in a modular fashion.
This means it can be easily expanded to accommodate different physical models and algorithms by implementing new classes which inherit from the class \verb|BaseHamiltonian| and \verb|BaseSolver|, respectively.
The functions and definitions necessary for the simulation of $2+1D$ free Wilson fermions have been implemented in the class \verb|FreeWilson2D|. This class also performs the explicit insertion of Eq.~\eqref{eq:qc-jwt-phi} and Eq.~\eqref{eq:qc-jwt-phi-dagger} into Eq.~\eqref{eq:num-h-free-tot-final-app}.

The ED algorithm has been implemented in the class \verb|ED|. This implementation uses the sparse linear algebra submodule of \verb|scipy|~\cite{scipy:2020}.
The VQE algorithm has been implemented in the class \verb|VQE|. For the optimizers we use the implementations provided by \verb|scipy|~\cite{scipy:2020}.
The ADAPT-VQE algorithm has been implemented in the class \verb|AdaptVQE|.
The implementation further utilizes the following python packages: \verb|Qiskit|~\cite{qiskit:2024}, \verb|scipy|~\cite{scipy:2020}, \verb|numpy|~\cite{numpy:2020}, \verb|sympy|~\cite{sympy:2017} and \verb|h5py|~\cite{h5py}.
In addition, the plots were created using \verb|matplotlib|~\cite{matplotlib:2007}.

\bibliography{references}
\end{document}